\documentclass[12pt,a4paper]{article}
\usepackage[preprint]{dfttrob}
\usepackage{times}
\usepackage[fleqn]{amsmath}
\usepackage{graphicx}
\usepackage{amssymb}
\dfttnum{DFTT 08/2002}

\newcommand{\pt}{\ensuremath{p_{\mathrm{t}}}}

\newcommand{\ee}{\mbox{e$^+$e$^-$}}

\newcommand {\mass} {\mbox{\rm GeV$\kern-0.15em /\kern-0.12em c^2$}}
\newcommand {\tev} {\mbox{${\rm TeV}$}}
\newcommand {\gev} {\mbox{${\rm GeV}$}}

\newcommand {\mom} {\mbox{\rm GeV$\kern-0.15em /\kern-0.12em c$}}

\newcommand {\gmom} {\mbox{\rm GeV$\kern-0.15em /\kern-0.12em c$}}
\newcommand {\mmass} {\mbox{\rm MeV$\kern-0.15em /\kern-0.12em c^2$}}
\newcommand {\mmom} {\mbox{\rm MeV$\kern-0.15em /\kern-0.12em c$}}

\newcommand{\K}{\mbox{$\mathrm {K}$}}

\newcommand{\Jpsi} {\mbox{J\kern-0.05em /\kern-0.05em$\psi$}\xspace}

\newcommand{\ppbar}{\mbox{$\mathrm {p\overline{p}}$}}

\newlength{\digitwidth} 
\newlength{\ql} \newlength{\qll} \newlength{\qlll}  \newlength{\qllll}
\newlength{\qlp} \newlength{\qllp} \newlength{\qlllp} \newlength{\qllllp}
\newlength{\qlup} \newlength{\qlupp}
\begin{document}

\providecommand{\SpS}{Sp\=pS}
\providecommand{\avg}[1]{\langle #1 \rangle}

\title{Correlations in proton-proton collisions with ALICE}
\author{A. Giovannini and R. Ugoccioni\\
Dipartimento di Fisica Teorica and I.N.F.N. -- sezione di Torino\\
via P. Giuria 1, 10125 Torino, Italy}
\maketitle
\begin{abstract}
Particle correlations and particle multiplicity distributions cannot be 
approached independently: a unified description of correlations and 
multiplicity distributions is always needed in order to understand the 
underlying dynamics in high energy collisions. In this light, we review
the most recent and interesting results on 
rapidity and momentum correlations,
emphasising the possibilities of measurements with the
ALICE detector.
\end{abstract}
\vfill
\begin{center}
Contribution to the ALICE Physics Performance Report, to appear.
\end{center}
\vspace*{2cm}

\newpage

\section{Particle correlations in pseudo-rapidity}
\label{TH:Particle_correlations_in_pseudo-rapidity}
Particle correlations and particle multiplicity distributions (MD's) cannot be 
approached independently:  a unified description of correlations and 
multiplicity distributions is always needed in order to understand the 
underlying dynamics in high energy collisions. The warning comes from
charged particle $H_q$  moments oscillations and shoulder structure in 
charged particle multiplicity distributions both in \ppbar\ collisions and
\ee\ annihilation, which have been shown to have a common origin in 
the weighted superposition of different classes of events or topologies, each 
one described by the same multiplicity distribution, i.e, a negative
binomial (NB) (Pascal) MD with characteristic parameters for each substructure 
\cite{TH:hqlett:2,TH:L3:mangeol}.
Accordingly, this Section should be considered the natural continuation of
Ref.~\cite{TH:Global_event_properties} on global event properties.

Attention here  will be   on  differential variables whose integrated  
versions   have been already discussed in the  just mentioned Reference.
Indeed, $n$-particle exclusive cross-section density   
\begin{equation}
	\frac{1}{\sigma_{\text{inel}}} \frac{d^n\sigma_n}{d\eta_1\dots d\eta_n}
\end{equation}
integrated over $\eta_1,\dots,\eta_n$, corresponds to  
$n$-particle multiplicity distribution 
$P_n = \sigma_n/ \sigma_{\text{inel}}$ , 
with $\sigma_n$ the $n$-particle topological 
cross-section and $\sum_n \sigma_n  = \sigma_{\text{inel}}$;
and $n$-particle inclusive cross-section density
\begin{equation}
  \frac{1}{\sigma_{\text{inel}}} \frac{d^n \sigma}{d\eta_1 ...d\eta_n}
   = \rho_n(\eta_1...\eta_n) 
\end{equation}
integrated over $\eta_1,\dots,\eta_n$, 
corresponds to $n$-th order factorial moments, $F_n$, of 
the multiplicity distribution $P_n$. 
In order to avoid inessential contributions
due to inclusive lower order combinatorics, 
the $n$-particle correlation functions
$C_n(\eta_1,\dots,\eta_n)$ are introduced. 
They are linked,
via standard cluster expansion of statistical mechanics,
to  $\rho_n(\eta_1...\eta_n)$.
Integrals of $C_n(\eta_1,\dots,\eta_n)$ 
over $\eta_1\dots\eta_n$ define factorial cumulant
moments, $K_n$, of the multiplicity distribution $P_n$. 
All the above mentioned
variables are not normalised, their corresponding normalised version can
be obtained by dividing by the
product of the single inclusive cross-sections.

It should be pointed out that  $C_n(\eta_1\dots\eta_n) = 0$ indicates
lack of $n$-particle correlations, a positive 
$C_n(\eta_1\dots\eta_n)$ the tendency
of the $n$ particles to be correlated and to group together  and 
a negative $C_n(\eta_1\dots\eta_n)$  
anti-correlations among particles. They prefer to stay far apart 
one from the other.

Moreover the description  of  different classes of 
events  in hadron-hadron collisions by 
NB (Pascal) MD's would suggest two-particle correlation dominance in rapidity
intervals as required by the hierarchical nature of $n$-particle correlations 
in each class of events \cite{TH:Void}.

\section{Two-particle pseudo-rapidity correlations}
\label{TH:Two-particle_pseudo-rapidity_correlations}
\begin{figure}
  \begin{center}
  \mbox{\includegraphics[width=0.9\textwidth]{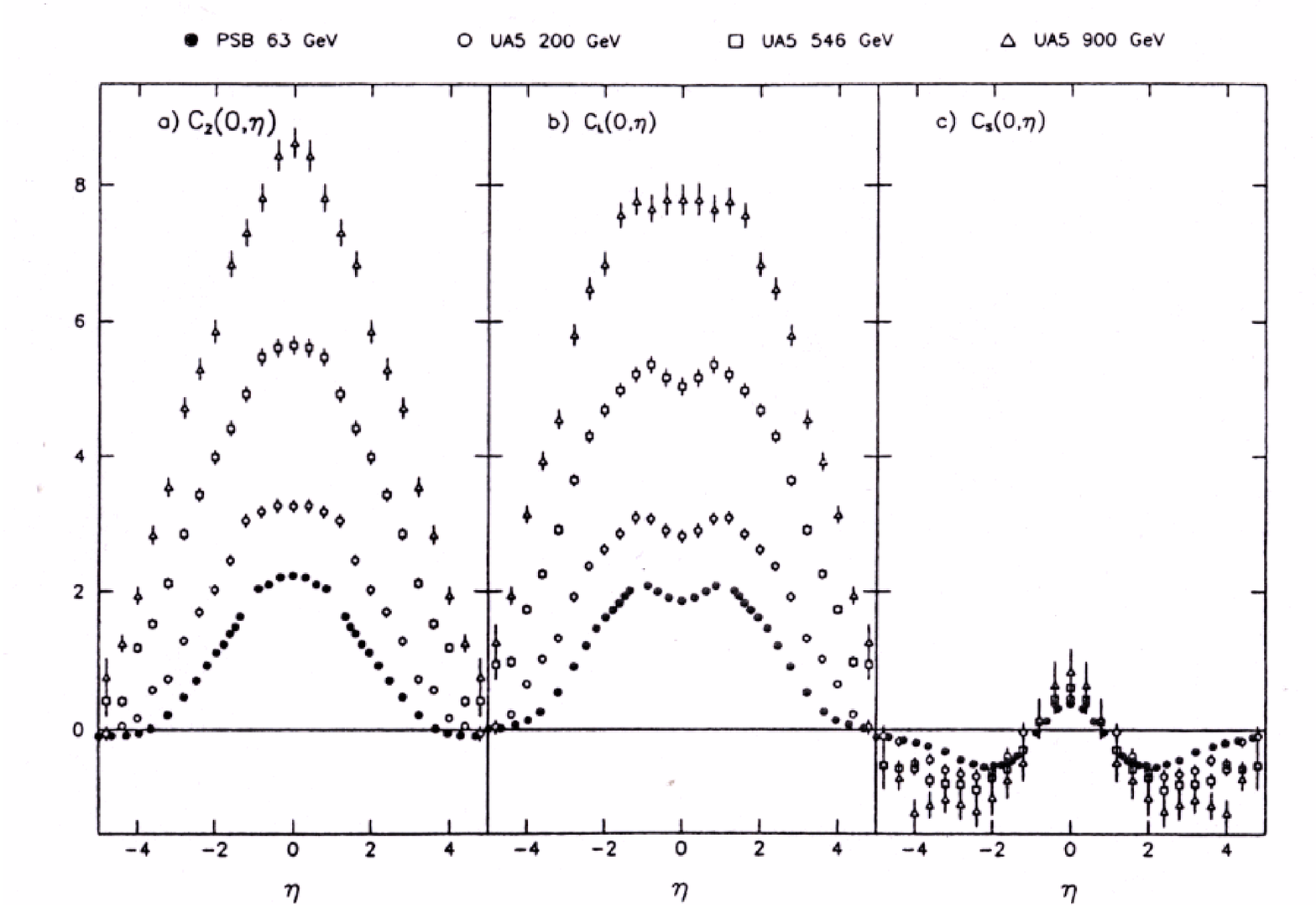}}
	%\vspace{3cm}
  \end{center}
  \caption{$C_2$,$C_L$ and $C_S$ at various c.m.\ 
	energies \cite{TH:DeWolf:rep}}\label{TH:fig:corr:1}
  \end{figure}

Two-particle pseudo-rapidity correlations
\begin{equation}
C_2 (\eta_1,\eta_2) = \rho_2 (\eta_1,\eta_2) - \rho_1(\eta_1)\rho_1(\eta_2) 
\end{equation}
have been shown to be strong  at small rapidity distances 
and to depend on the particles' charge combination.
It should be reminded that by integrating $C_2( \eta_1 ,\eta_2)$ over
$\eta_1$ and $\eta_2$ one gets $D^2 - \bar n$, where $D$ is the dispersion.
In hadron-hadron collisions, the energy dependence of
$C_2(\eta_1=0,\eta )$ from ISR 
\cite{TH:ISR:correlations} to UA5 \cite{TH:UA5:correlations} energies 
is described in Fig.~\ref{TH:fig:corr:1}a.

\begin{figure}
  \begin{center}
  \mbox{\includegraphics[width=0.4\textwidth]{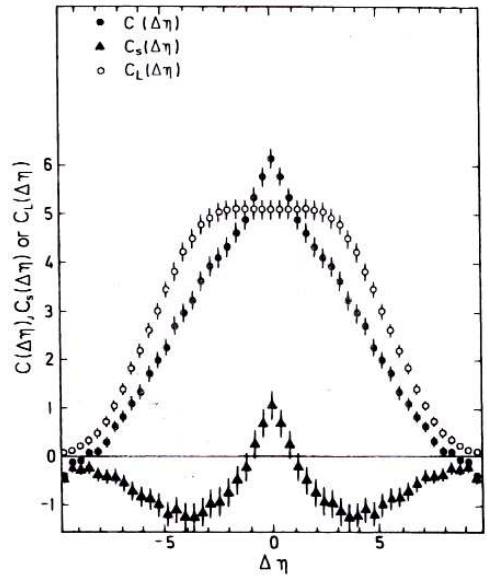}}	
	%\vspace{3cm}
  \end{center}
  \caption{Same as previous figure but only 546~\gev\ is 
	shown. \cite{TH:UA5:rep}}\label{TH:fig:corr:2}
  \end{figure}

Inclusive two-particle correlation functions $C_2(\eta_1,\eta_2)$
can be written in terms of semi-inclusive single- and two-particle 
densities for particles $a$ and $b$:
\begin{equation}
 \rho^{(n)}_1 (\eta) = \frac{1}{\sigma_n}  \frac{d^n \sigma^a_n}{d \eta}
\end{equation}
and
\begin{equation}
 \rho^{(n)}_2 (\eta_1, \eta_2) = 
			\frac{1}{\sigma_n} \frac{d^n \sigma^{ab}_n}{d \eta_1 d\eta_2}
\end{equation}
by mixing events of different  multiplicities $n$:
two new functions can be defined:
\begin{equation}
  C_2 (\eta_1,\eta_2) = C_S (\eta_1,\eta_2) + C_L (\eta_1,\eta_2)  
\end{equation}
where 
\begin{equation}
C_S (\eta_1,\eta_2) =  \sum_n P_n C^{(n)}_2(\eta_1,\eta_2) 
\end{equation}
and 
\begin{equation}
C_L (\eta_1,\eta_2) =  
  \sum_n P_n [\rho^{(n)}_1(\eta_1)-\rho_1(\eta_1)]
- \sum_n P_n [\rho^{(n)}_1(\eta_2)-\rho_1(\eta_2)]
\end{equation}
with  
\begin{equation}
  C^{(n)}_2 (\eta_1,\eta_2)= \rho^{(n)}_2(\eta_1,\eta_2)
    - \rho^{(n)}_1 (\eta_1) \rho^{(n)}_1 (\eta_2)
\end{equation}
i.e., the two-particle correlation function at fixed multiplicity $n$.

$C_S$ is the average of the semi-inclusive correlation functions
and is more sensitive to dynamical correlations, and $C_L$ describes  the
mixing of different topological single particle densities.
As shown in Fig.~\ref{TH:fig:corr:1},  
$C_S$  does not depend significantly on energy.
This fact should be contrasted with $C_L$ behaviour which is strongly 
energy dependent  and is the result  of mixing different 
topological single particle densities. In Fig.~\ref{TH:fig:corr:1}b
and \ref{TH:fig:corr:1}c are shown the 
energy dependence of $C_S$ and $C_L$: the contribution of $C_L$
to $C_2( \eta_1,\eta_2)$
is dominant with respect to $C_S$, in addition it increases significantly
 with energy whereas $C_S$ does not grow as much with energy around $\eta=0$:
in addition $C_S$ is positive in the just mentioned region and becomes
negative for $|\eta| > 1$.
The overall behaviour of $C_2$, $C_L$ and $C_S$  at fixed c.m.\ energy is
summarised in Fig.~\ref{TH:fig:corr:2}.

It should be pointed out that
\begin{equation}
  \int C_S (\eta_1,\eta_2) d \eta_1 d\eta_2 = - \bar n 
\end{equation}
and
\begin{equation}
  \int C_L (\eta_1,\eta_2) d \eta_1 d\eta_2 = D^2.
\end{equation}
These equations clarify the meaning of the two contributions to total
two particle correlations $C_2 (\eta_1,\eta_2)$ and point out
what quantity should be measured with ALICE in the allowed pseudo-rapidity
range.   

\section{Bose-Einstein correlations}
\label{TH:Bose-Einstein_correlations}

Among two-particle correlations, Bose-Einstein (BE)
correlations are of particular
interest in high energy collisions.
The production of two identical bosons 
$a$, $b$ from two particle
sources is controlled by an amplitude which is symmetrised  with respect
to the interchange  of bosons $a$, $b$  and leads to an enhanced 
emission probability if bosons have similar momenta.

BE correlations are measured in terms of 
the second order normalised factorial  cumulant $R_2$, i.e.,
\begin{equation}
	R_2(q_a,q_b) = \frac{\rho_2(q_a,q_b)}{\rho_1(q_a)\rho_1(q_b)} - 1  ,
\end{equation}
the ratio of two-particle inclusive cross-section 
$\sigma^{-1}d^2\sigma/d q_a d q_b$ 
over the product of single- $a$ and $b$ particle inclusive cross-sections 
$\sigma^{-1}d\sigma/d q_a$, $\sigma^{-1}d\sigma/ d q_b$.

$R_2$ function is directly related to the Fourier transform of 
the  space-time distribution of particle production points.
Accordingly, space distribution and lifetime of boson sources can be
measured.  In the standard Goldhaber parametrisation  
\begin{equation}
	 R_2 (Q) = \lambda \exp\left( - G^2 Q^2 \right)  ,
\end{equation}
with $Q^2 = - (q_a - q_b)^2$,   the square of the four momentum difference
of particles $a$ and $b$, [$q_{a,b} \equiv (\vec{p}_{a,b},E_{a,b})$];
$\lambda$ is the strength of the effect and $G$ is a measure of the
source size.

The effect depends strongly on the masses  of particles used in the
analysis. It is  well known  that the range  of the radius of the
source from which pions are produced in \ee\ annihilation at
LEP lies between $\approx\,0.7$-$1.0$~fm, the range of the radius of the
source from which \K\ particles are emitted between  $\approx\,0.5$-$0.7$~fm
and that for $\Lambda$ particles between $\approx\,0.1$-$0.2$~fm.
In conclusion
the most external shell produces particle pairs of lower mass (pions)
and higher mass particles pairs (lambdas) are coming from most internal 
shells.

\begin{figure}
  \begin{center}
  \mbox{\includegraphics[width=0.48\textwidth]{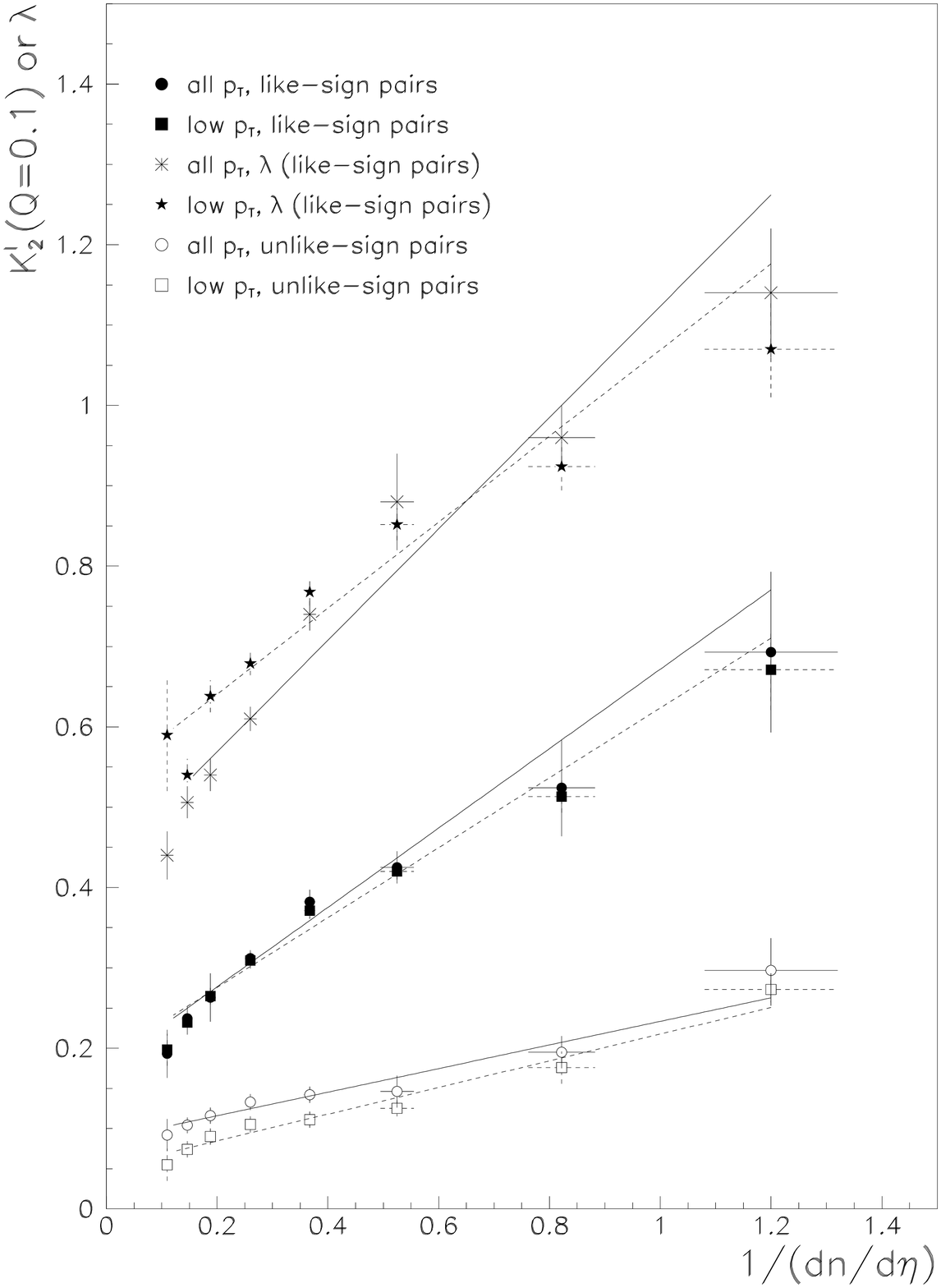}}
  \mbox{\includegraphics[width=0.48\textwidth]{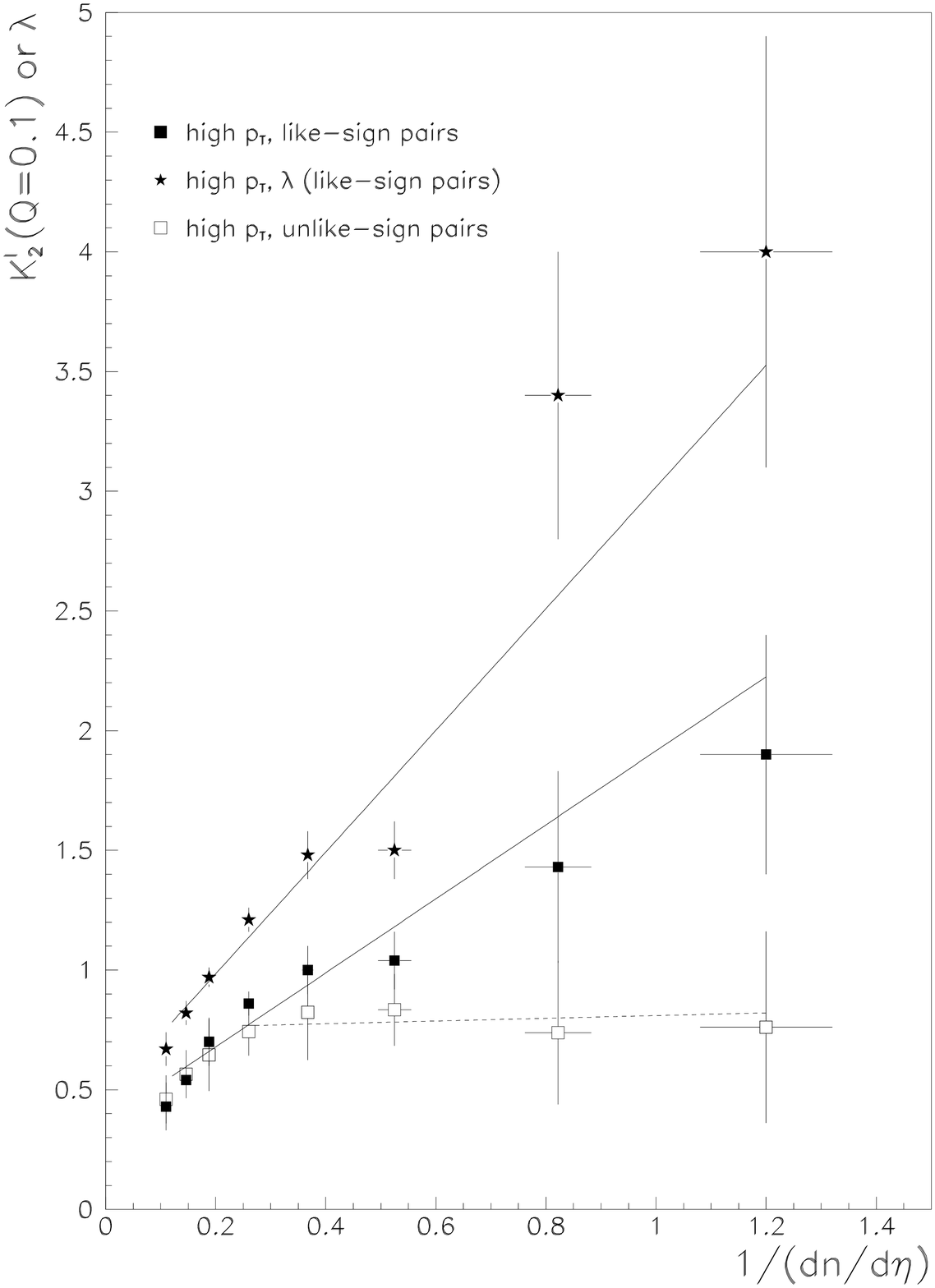}}
  \end{center}
  \caption{Analysis of UA1 data \cite{TH:Buschbeck:Torino2000}: 
   the intercept of the BE correlation
   (minus 1) is plotted against the inverse of the multiplicity
   density separately for high-\pt\ ($\pt > 0.7$ \gmom) and low-\pt\
	particles}\label{TH:fig:Buschbeck_HBT}
  \end{figure}

Are  different particle production shells visible also in pp collisions 
in the \tev\ region? 
does a connection exist between  low pair mass scale and short distance  
scale of QCD mini-jets?
Extremely high ALICE statistics should allow to clarify all these points.
Bose-Einstein effect depends also on \pt\ intervals considered and on related 
multiplicity densities \cite{TH:buschbeck:2000,TH:Buschbeck:Torino2000}.
A sample of 2,400,000 non-single-diffractive \ppbar\ events
at c.m.\ energy 630~\gev\  measured by the UA1 central
detector have been used. Only vertex associated charged tracks with
$\pt > 0.15$~\gmom, and $|\eta| < 3$ have been considered. The azimuth angle
has been restricted to $45^{\text{o}} < |\phi| < 135^{\text{o}}$.

Three subsamples for like-sign (ls) and opposite-sign (os) pion pairs
have been selected  with respect to  pion transverse momentum \pt:
\begin{itemize}
\item[i.] the all-\pt\ sample;
\item[ii.] a low-\pt\ subsample of charged particles 
with $\pt < 0.7$~\gmom\ (attention is 
paid in order to reduce the number of particles originated from jets or 
mini-jets);
\item[iii.] a high-\pt\ subsample  with $p_T > 0.7$~\gmom\ 
(particles come here predominantly from 
jets or mini-jets).
\end{itemize}

Then the behaviour of second order normalised cumulant correlation 
functions has been studied in the three subsamples for
particle multiplicity densities $dN/d\eta = 1.22$, $2.72$ and $6.85$.
BE correlations show in hadron-hadron collisions  a
pronounced dependence on multiplicity  in the BE strength parameter
lambda or equivalently the cumulant of second order $K_2$ at $Q=0$.
In particular the high-\pt\ sample cumulants exceeds one  (which would
indicate full coherence in pure BE correlations) and is  hence
concluded that this data sample is dominated by processes other than
BE correlations (Fig.~\ref{TH:fig:Buschbeck_HBT}).

As far as the low-\pt\ sample is concerned it is shown  that the dependence
of  the correlation strength  and of higher order cumulants on
multiplicity is important in order to test different theoretical models
(Monte Carlos are totally inadequate here): for this task the low \pt\
cut-off of ALICE at LHC is required.

\section{Higher moment correlations in small pseudo-rapidity intervals}
\label{TH:Higher_moment_correlations}

The presence of very short range correlations and its connection
with BE interference is indeed an important topic to be investigated,
as in more general terms it is the search of local fluctuations
of multiplicity distributions  in momentum space and related scaling
properties.  Large concentrations of particle number in
small pseudo-rapidity intervals for single events have been seen  in the
JACEE event for a pseudo-rapidity binning $\delta\eta = 0.1$ 
(local fluctuations up to  $dn/d\eta$  $\approx\,300$ with a
signal-to-background ratio about 1:1)
\cite{TH:Inter:1}
and in the NA22 event (local fluctuation in rapidity of 60 times the
average density)
\cite{TH:Inter:2}.

\begin{figure}
  \begin{center}
  \mbox{\includegraphics[width=0.6\textwidth]{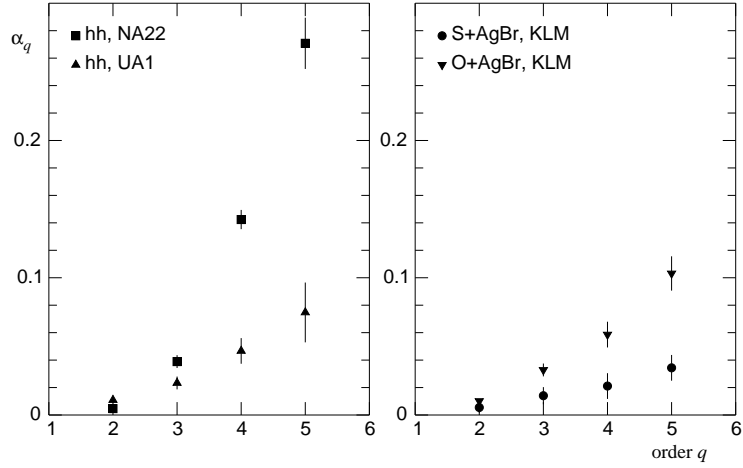}}
  \end{center}
  \caption{Intermittency slope in different reactions, comparing
  the behaviour in hadron-hadron collisions to that in nucleus-nucleus ones 
	\cite{TH:Bialas:1991gt}.}\label{TH:fig:intermittency}
  \end{figure}

A possible explanation of these spikes was related to an intermittent
behaviour, i.e., to the guess that  there exist a correlations at all scales  
which implies a power low  dependence of normalised factorial moments 
$F_q (\delta)$ on the size $\delta$ of the phase space bins:
\begin{equation}
  \frac{F_q (\delta)}{\avg{n(\delta)}^q}  =  \delta^{-\alpha_q} ;
\end{equation}
the slope $\alpha_q$ was shown to increase as $q$ increases, see
Fig.~\ref{TH:fig:intermittency}.

Since ALICE allows to study with its high statistics  moments properties
up to order $q=14$, it will be particularly interesting 
to analyse both  intermittent 
behaviour  as bin size decreases and the influence on this phenomenon on
BE correlations together with resonance production.
  
If the above mentioned  scaling law will be  confirmed in the LHC energy
 domain a new horizon  will be open on self-similar cascading structure
and fractal properties of  hadron hadron collisions, a fascinating
perspective also for more complex collisions, like heavy ions.

\section*{Acknowledgements}
We thank J.-P. Revol for stimulating the present work.


\section*{References}
\begin{thebibliography}{110}
\parskip=0.pt \parsep=0.pt \itemsep=0.pt

\bibitem{TH:Global_event_properties}
{A. Giovannini and R. Ugoccioni}, University of Torino preprint DFTT 29/2001.

\bibitem{TH:hqlett:2}
{A.~Giovannini, S.~Lupia and R.~Ugoccioni}, Phys.\ Lett.\ {\bf B374}  (1996)
  231.

\bibitem{TH:L3:mangeol}
{P. Achard et al.\ (L3 Collaboration)}, preprint CERN-EP/2001-072
  (hep-ex/0110072), CERN.

\bibitem{TH:Void}
{S. Lupia, A. Giovannini and R. Ugoccioni}, Z. Phys.\ {\bf C59}  (1993) 427.

\bibitem{TH:ISR:correlations}
{Amendolia et al.}, Nuovo Cimento {\bf 31A}  (1976) 19.

\bibitem{TH:UA5:correlations}
{R.E.\ Ansorge et al.\ (UA5 Collaboration)}, Z. Phys.\ {\bf C37}  (1988) 191.

\bibitem{TH:DeWolf:rep}
{E.A.~De Wolf, I.M.~Dremin and W.~Kittel}, Physics Reports {\bf 270}  (1996) 1.

\bibitem{TH:UA5:rep}
{G.J.\ Alner et al.\ (UA5 Collaboration)}, Physics Reports {\bf 154}  (1987)
  247.

\bibitem{TH:Buschbeck:Torino2000}
{B. Buschbeck and H.C. Eggers}, Nucl.\ Phys.\ (Proc.\ Suppl.) {\bf B92}  (2001)
  235.

\bibitem{TH:buschbeck:2000}
{B. Buschbeck, H.C. Eggers, and P. Lipa}, Phys.\ Lett.\ {\bf B481}  (2000) 187.

\bibitem{TH:Inter:1}
{T.H. Burnett et al.\ (JACEE Collaboration)}, Phys.\ Rev.\ Lett.\ {\bf 50}
  (1983) 2062.

\bibitem{TH:Inter:2}
{M. Adamus et al.\ (NA22 Collaboration)}, Phys.\ Lett.\ {\bf B185}  (1987) 200.

\bibitem{TH:Bialas:1991gt}
{A. Bia{\l}as}, NuclPhys {\bf A525}  (1991) 345.


\end{thebibliography}
\end{document}